\documentclass[a4paper,11pt]{article}

\usepackage{a4wide,amsmath,natbib,graphicx,url}
\usepackage[french,english]{babel}
\bibpunct{(}{)}{;}{a}{,}{,}

\newcommand{\PP}{\ensuremath{\operatorname{P}}}
\newcommand{\smbeta}{\ensuremath{\lambda_{\beta}}}
\newcommand{\smalpha}{\ensuremath{\lambda_{\alpha}}}
\newcommand{\nd}{\ensuremath{{n_D}}}
\newcommand{\nt}{\ensuremath{{n_T}}}

\newcommand{\nw}{\ensuremath{{n_W}}}

\begin{document}
\selectlanguage{english}

\title{Inference and Evaluation of the Multinomial Mixture Model\\
  for Text Clustering}
\author{Lo\"{\i}s Rigouste, Olivier Capp\'e and Fran\c{c}ois Yvon\\
  (rigouste, cappe, yvon) at enst.fr \\
  GET / T\'el\'ecom Paris \& CNRS / LTCI~\footnote{This work has been supported
    by France T\'{e}l\'{e}com, Division R\&D, under contract
    n$^\circ$42541441.}\\
  46 rue Barrault, 75634 Paris c\'{e}dex 13, France}
\date{} 

\maketitle

\begin{abstract}
  In this article, we investigate the use of a probabilistic model for
  unsupervised clustering in text collections. Unsupervised clustering has
  become a basic module for many intelligent text processing applications, such
  as information retrieval, text classification or information extraction.

  Recent proposals have been made of probabilistic clustering models, which
  build ``soft'' theme-document associations. These models allow to compute, for
  each document, a probability vector whose values can be interpreted as the
  strength of the association between documents and clusters. As such, these
  vectors can also serve to project texts into a lower-dimensional ``semantic''
  space. These models however pose non-trivial estimation problems, which are
  aggravated by the very high dimensionality of the parameter space.

  The model considered in this contribution consists of a mixture of
  multinomial distributions over the word counts, each component corresponding
  to a different theme. We present and contrast various estimation procedures,
  which apply both in supervised and unsupervised contexts. In supervised
  learning, this work suggests a criterion for evaluating the posterior odds of
  new documents which is more statistically sound than the ``naive Bayes''
  approach. In an unsupervised context, we propose measures to set up a
  systematic evaluation framework and start with examining the
  Expectation-Maximization (EM) algorithm as the basic tool for inference. We
  discuss the importance of initialization and the influence of other features
  such as the smoothing strategy or the size of the vocabulary, thereby
  illustrating the difficulties incurred by the high dimensionality of the
  parameter space. We also propose a heuristic algorithm based on iterative EM
  with vocabulary reduction to solve this problem. Using the fact that the
  latent variables can be analytically integrated out, we finally show that
  Gibbs sampling algorithm is tractable and compares favorably to the basic
  expectation maximization approach.

  \textsl{Keywords}: Multinomial Mixture Model, Expectation-Maximization, Gibbs Sampling,
  Text Clustering
\end{abstract}

\selectlanguage{french}
\begin{abstract}
  Dans cet article, nous pr\'esentons une \'etude d\'etaill\'ee d'un mod\`ele
  probabiliste simple, le m\'elange de multinomiales, dans un contexte de
  classification non-supervis\'ee de collections de textes.

  La construction de groupes de documents th\'ematiquement homog\`enes est une
  des technologies de base de la fouille de texte, et trouve de multiples
  applications, aussi bien en recherche documentaire qu'en cat\'egorisation de
  documents, ou encore pour le suivi de th\`emes et la construction de
  r\'esum\'es. Diverses propositions r\'ecentes ont \'et\'e faites de mod\`eles
  probabilistes permettant de construire de tels regroupements. Les mod\`eles de
  classification probabiliste ont l'avantage de pouvoir \'egalement \^etre vus
  comme des outils permettant de construire des repr\'esentations num\'eriques
  synth\'etiques des informations contenues dans le document.  Ces mod\`eles posent
  toutefois des probl\`emes d'estimation difficiles, qui sont 
  d\^us en particulier \`a la tr\`es grande dimensionalit\'e du vocabulaire.

  Notre contribution \`a cette famille de travaux est double: nous pr\'esentons
  d'une part plusieurs algorithmes d'inf\'erence, certains originaux, pour
  l'estimation du mod\`ele de m\'elange de multinomiales; nous pr\'esentons
  \'egalement une \'etude syst\'ematique des performances de ces algorithmes,
  fournissant ainsi de nouveaux outils m\'ethodologiques pour mesurer les
  performances des outils de classification non supervis\'ee.
\end{abstract}

\selectlanguage{english}
\section{Introduction}
\label{sec:intro}

The wide availability of huge collections of text documents (news corpora,
e-mails, web pages, scientific articles...) has fostered the need for efficient
text mining tools. Information retrieval, text filtering and classification, and
information extraction technologies are rapidly becoming key components of
modern information processing systems, helping end-users to select, visualize
and shape their informational environment.

Information retrieval technologies seek to rank documents according to their
relevance with respect to users queries, or more generally to users
informational needs. Filtering and routing technologies have the potential to
automatically dispatch documents to the appropriate reader, to arrange incoming
documents in the proper folder or directory, possibly rejecting undesirable
entries. Information extraction technologies, including automatic summarization
techniques, have the additional potential to reduce the burden of a full reading
of texts or messages. Most of these applications take advantage of
(unsupervised) \emph{clustering techniques} of documents or of document
fragments: the unsupervised structuring of documents collections can for
instance facilitate its indexing or search; clustering a set of documents in
response to a user query can greatly ease its visualization; considering
sub-classes induced in a non-supervised fashion can also improve text
classification \citep{vinot03improving}, etc. Tools for building
thematically coherent sets of documents are thus emerging as a basic
technological block of an increasing number of text processing applications.

Text clustering tools are easily conceived if one adopts, as is commonly done, a
\emph{bag-of-word} representation of documents: under this view, each text is
represented as a high-dimensional vector which merely stores the counts of each
word in the document, or a transform thereof. Once documents are turned into
such kind of numerical representation, a large number of clustering techniques
become available \citep{jain99clustering} which allow to group documents based
on ``semantic'' or ``thematic'' similarity. For text clustering tasks, a number
of proposal have recently been made which aim at identifying probabilistic
(``soft'') theme-document associations (see,
eg., \citealp{hofmann01unsupervised,blei02latent,buntine04discretepca}). These probabilistic
clustering techniques compute, for each document, a probability vector whose
values can be interpreted as the strength of the association between documents
and clusters. As such, these vectors can also serve to project texts into a
lower-dimensional space, whose dimension is the number of clusters.  These
probabilistic approaches are certainly appealing, as the projections they build
have a clear, probabilistic interpretation; this is in sharp contrast with
alternative projection techniques for text documents, such as Latent Semantic
Analysis (LSA) \citep{deerwester90lsa} or non-negative matrix factorization
(NMF) techniques \citep{vinokourov02why,shanaz06document}.

In this paper, we focus on a simpler probabilistic model, in which the corpus is
represented by a mixture of multinomial distributions, each component
corresponding to a different ``theme''
\citep{nigam00text}. This model is the unsupervised
counterpart of the popular ``Naive Bayes'' model for text classification (see,
eg., \citealp{lewis98naivebayes,mccallum98naivebayes}). Our main objective is to
analyze the estimation procedures that can be used to infer the model parameters,
and to understand precisely the behavior of these estimation procedures when
faced with high-dimensional parameter spaces. This situation is typical of the
bag-of-word model of text documents but may certainly occur in other contexts
(bioinformatics, image processing\ldots). Our contribution is thus twofold:
\begin{itemize}
\item we present a comprehensive review of the model and of the estimation
  procedures that are associated with this model, and introduce novel variants
  thereof, which seem to yield better estimates for high-dimensional models, and
  report a detailed experimental analysis of their performance.
\item these analyses are supported by a methodological contribution on the
  delicate, and often overlooked, issue of performance evaluation of clustering
  algorithms (see, eg., \citealp{halkidi01clustering}). Our proposal here is to
  focus on a ``pure'' clustering tasks, where the number of themes (the number
  of dimensions in the ``semantic'' space) is limited, which allows in our case
  a direct comparison with a reference (manual) clustering.
\end{itemize}

This article is organized as follows. We firstly introduce the model and notations
used throughout the paper. Dirichlet priors are set on the parameters and we may
use the Expectation-Maximization (EM) algorithm to obtain \emph{Maximum A
  Posteriori} (MAP) estimates of the parameters. An alternative inference
strategy uses simulation techniques (Monte-Carlo Markov Chains) and consists in
identifying conditional distributions from which to generate samples. We show,
in Section~\ref{sec:loi_cond}, that it is possible to marginalize analytically
all continuous parameters (thematic probabilities and theme-specific word
probabilities). This result generalizes an observation that was used, in the
context of the Latent Dirichlet Allocation (LDA) model by
\citep{griffiths02probabilistic}. We first examine what the consequences of this
derivation are for supervised classification tasks. We then describe our
evaluation framework and highlight, in a first round of experiments, the
importance of the initialization step in the EM algorithm. Looking for ways to
overstep the limitations of EM by incremental learning, we present an algorithm
based on a progressive inclusion of the vocabulary. We eventually discuss the
application of Gibbs sampling to this model, reporting experiments which
support the claim that, in our context, the sampling based approach is more
robust than EM alternatives.

\section{Basics}
\label{sec:model}

In this section, we present our model of the count vectors. Since we
assume that the distribution of the words in the document depends on
the value of a latent variable associated with each text, the
\emph{theme}, we use a multinomial mixture model with Dirichlet priors
on the parameters.

We show how this model is related to the naive Bayes classifier and
then explain that some conditional densities follow another
distribution, called ``Dirichlet-Multinomial'' and how this fact
proves useful for both classification and unsupervised learning.

\subsection{Multinomial Mixture Model}
We denote by $\nd$, $\nw$ and $\nt$, respectively, the number of
documents, the size of the vocabulary and the number of themes, that
is, the number of components in the mixture model. Since we use a
bag-of-words representation of documents, the corpus is fully
determined by the count matrix $C=(C_{wd})_{w=1\dots{}\nw,
  d=1\dots{}\nd}$; the notation $C_d$ is used to refer to the word
count vector of a specific document $d$. The multinomial mixture model is such that:
\begin{equation}
  \PP(C_d|\alpha,\beta) = \sum_{t=1}^{\nt}\alpha_t \, \frac{l_d!}{\prod_{w=1}^{\nw}C_{wd}!}\prod_{w=1}^{\nw}\beta_{wt}^{C_{wd}}
\end{equation}
Note that the document length itself (denoted by $l_d$) is taken as an
exogenous variable and its distribution is not accounted for in the model.  The
notations $\alpha = (\alpha_1,\alpha_2,\ldots,\alpha_{\nt})$ and $\beta_{t} =
(\beta_{1t},\beta_{2t},\ldots,\beta_{{\nw}t})$ (for $t=1,\ldots,\nt$) are used
to refer to the model parameters, respectively, the mixture weights and the
collection of theme-specific word probabilities.

Adopting a Bayesian approach, we set independent noninformative Dirichlet
priors on $\alpha$ (with hyperparameter $\smalpha>0$) and on the columns
$\beta_{t}$ (with hyperparameter $\smbeta>0$). The choice of the Dirichlet
distribution in this context is natural because it is the conjugated
distribution associated to the multinomial, a property which will be
instrumental in Section~\ref{sec:loi_cond}.

Therefore we get the following probabilistic generative mechanism for
the whole corpus $C=(C_1\ldots{}C_{\nd})$:
\begin{enumerate}
\item sample $\alpha$ from a Dirichlet distribution with probabilities
  $\smalpha,\ldots,\smalpha$
\item for every theme $t=1,\ldots,\nt$, sample $\beta_{t}$ from a Dirichlet distribution with probabilities $\smbeta,\ldots,\smbeta$
\item
for every document $d=1,\ldots,\nd$
\begin{enumerate}
\item sample a theme $T_d$ in $\{1,\dots{},\nt\}$ with probabilities
  $\alpha = (\alpha_1,\alpha_2,\ldots,\alpha_{\nt})$
\item sample $l_d$ words from a multinomial distribution with theme-specific probability vector $\beta_{T_d}$.
\end{enumerate}
\end{enumerate}

As all documents are assumed to be independent, the corpus
 likelihood is given by
\begin{eqnarray*}
\PP(C|\alpha,\beta) = \prod_{d=1}^\nd \PP (C_d|\alpha,\beta)
\end{eqnarray*}

Now, as the prior distributions are Dirichlet, the posterior
distribution is proportional to (disregarding terms that do not depend
on $\alpha$ or $\beta$):
\begin{eqnarray}
p(\alpha,\beta|C)&\propto&\PP(C|\alpha,\beta)p(\alpha)p(\beta)\nonumber\\
&\propto&\left(\prod_{d=1}^\nd\sum_{t=1}^{\nt}\alpha_t
  \prod_{w=1}^{\nw}\beta_{wt}^{C_{wd}}\right)\prod_{t=1}^\nt\alpha_t^{\smalpha-1}\prod_{t=1}^\nt\prod_{w=1}^\nw\beta_{wt}^{\smbeta-1}\label{eq:paramdist}
\end{eqnarray}
Maximizing this expression is in general intractable.

We first consider the simpler case of \emph{supervised inference} in which the
themes $T = (T_1, \dots, T_\nd)$ associated with the documents are observed. In
this situation, inference is based on $p(\alpha,\beta|C,T)$ rather than
$p(\alpha,\beta|C)$. In Section~\ref{sec:naivebayes}, we briefly recall that
maximizing $p(\alpha,\beta|C,T)$ with respect to $\alpha$ and $\beta$ yields
the so-called naive Bayes classifier (with Laplacian smoothing). In
section~\ref{sec:loi_cond}, we turn to the so-called fully Bayesian inference
which consists in integrating with respect to the posterior distribution
$p(\alpha,\beta|C,T)$. This second approach yields an alternative
classification rule for unlabeled documents which is connected to the
Dirichlet-Multinomial (or Polya) distribution. Both of these approaches have
counterparts in the context of unsupervised inference which will be developed
in Sections~\ref{ssec:em}. and~\ref{ssec:gibbs}, respectively.

\subsection{Naive Bayes Classifier}
\label{sec:naivebayes}
When $T$ is observed, the log-posterior distribution of the parameters
given both the documents $C$ and their themes $T$ has the simple form:
\begin{equation}
\log p(\alpha,\beta|C,T) = \sum_{t=1}^{\nt}\left( (S_t+\smalpha-1) \log\alpha_t + \sum_{w=1}^{\nw}(K_{wt}+\smbeta-1)\log\beta_{wt}\right)
\label{eq:logpost}
\end{equation}
up to terms that do not depend on the parameters, where $S_t$ is the number of
training documents in theme $t$ and $K_{wt}$ is the number of occurrences of
the word $w$ in theme $t$.

Taking into account the constraints $\sum_{t=1}^\nt\alpha_t = 1$ and
$\sum_{w=1}^\nw\beta_{wt} = 1$ (for $t\in\{1,\ldots,\nt\}$), the maximum a
posteriori estimates have the familiar form:
\begin{align*}
\hat{\alpha}_t = \frac{S_t+\smalpha-1}{\nd+\nt(\smalpha-1)} \qquad
\hat{\beta}_{wt} = \frac{K_{wt}+\smbeta-1}{K_t+\nw(\smbeta-1)}
\end{align*}
were $K_t=\sum_{w=1}^\nw K_{wt}$ is the total number of occurrences in theme
$t$.

In the following, we will denote quantities that pertain to a test
corpus distinct from the training corpus $C$ using the $\star$
superscript. Thus $C^\star$ is the test corpus, $C^\star_d$ a
particular document in the test corpus, $l_d^\star$ its length, etc.
The Bayes decision rule for classifying an unlabeled test document,
say $C^\star_d$, then consists in selecting the theme $t$ which
maximizes
\begin{align}
  \PP(T^\star_d =t | C^\star_d,\hat{\alpha},\hat{\beta}) & = \hat{\alpha_t}
  \prod_{w=1}^\nw \hat{\beta}_{wt}^{C_{wd}^\star} \nonumber \\
  & \propto (S_t+\smalpha-1) \frac{\prod_{w=1}^\nw
    (K_{wt}+\smbeta-1)^{C_{wd}^\star}}{(K_t+\nw(\smbeta-1))^{l_{d}^\star}}
  \label{eq:naivebayes}
\end{align}
The above formula corresponds to the so-called naive Bayes classifier, using
Laplacian smoothing for word and theme probability estimates
\citep{lewis98naivebayes,mccallum98naivebayes}.

\subsection{Fully Bayesian Classifier}
\label{sec:loi_cond}
An interesting feature of this model is that it is also possible to integrate
out the parameters $\alpha$ and $\beta$ under their posterior distribution
allowing to evaluate the Bayesian predictive distribution
\begin{equation}
  \label{eq:predictdist}
  \PP(T^\star_d=t|C^\star_d, C,T) = \int \PP(T^\star_d=t|C^\star_d,\alpha,\beta) p(\alpha,\beta|C,T) d \alpha d \beta
\end{equation}
From a Bayesian perspective, this predictive distribution is preferable, for
classifying the document $C^\star_d$, to the naive Bayes rule given
in~\eqref{eq:naivebayes}. Tractability of the above integral stems from the
fact that $p(\alpha,\beta|C,T)$ is a product of Dirichlet distributions --
see~\eqref{eq:logpost}. Hence $\PP(C^\star_d|T^\star_d=t)$ follows a so called
\emph{Dirichlet-Multinomial} distribution
\citep{mosimann62dirmult,minka03dirichlet}.

To see this, consider the joint distribution of the observations $C$, the latent variables $T$ and the parameters $\alpha$ and $\beta$:
\begin{equation*}
  \PP(C,T,\alpha,\beta)
    \propto \prod_{t=1}^{\nt} \left( \alpha_t^{S_t +
      \smalpha -1 } \prod_{w=1}^{\nw} \beta_{wt}^{K_{wt} + \smbeta -1} \right)
\end{equation*}
As the above quantity, viewed as a function of $\alpha$ and $\beta_1, \dots,
\beta_\nt$, is a product of unnormalized Dirichlet distributions, it is
possible to integrate out $\alpha$ and $\beta$ analytically. The result of
the integration involves the normalization constants of the Dirichlet
distributions, yielding:
\begin{eqnarray}
\PP(T|C) &\propto &\frac{\prod_{t=1}^\nt \Gamma(S_t +
  \smalpha)}{\Gamma\left[\sum_{t=1}^\nt \left(S_t + \smalpha \right)\right]}
\prod_{t=1}^\nt \frac{\prod_{w=1}^\nw
  \Gamma(K_{wt}+\smbeta)}{\Gamma\left[\sum_{w=1}^\nw (K_{wt} +
    \smbeta)\right]} \nonumber \\
&\propto &\prod_{t=1}^\nt \left(\Gamma(S_t +
  \smalpha)\frac{\prod_{w=1}^\nw
  \Gamma(K_{wt}+\smbeta)}{\Gamma\left[\sum_{w=1}^\nw (K_{wt} + \smbeta)\right]}\right)
\label{eq:jointconditional}
\end{eqnarray}

Now, if we single out the document of index $d$ assuming that the document
$C_d$ itself has been observed but that the theme $T_d$ is unknown, elementary
manipulations yield:
\begin{equation}
\PP(T_d = t |C_d, C_{-d}, T_{-d})
 \propto \left(S_{t} -1 + \smalpha \right)
\frac{\prod_{w=1}^\nw \Gamma(K_{wt}+\smbeta)}{\prod_{w=1}^\nw
  \Gamma(K_{wt}^{-d}+\smbeta)} \frac{\Gamma\left[\sum_{w=1}^{\nw} (K_{wt}^{-d} +
    \smbeta)\right]}{\Gamma\left[\sum_{w=1}^\nw (K_{wt} + \smbeta)\right]}
   \label{eq:loi_cond}
\end{equation}
where $T_{-d}$ is the vector of theme indicators for all documents but $d$,
$C_{-d}$ denotes the corpus deprived from document $d$, and $K_{wt}^{-d}$ is
the quantity $K_{wt}^{-d} = \sum_{\{d'\ne{}d:T_d = t\}} C_{wd}$. With suitable
notation change, this is exactly the predictive distribution as defined
in~\eqref{eq:predictdist}.

Note that, in contrast to the case of the joint posterior probabilities
$\PP(T|C)$ given in~\eqref{eq:jointconditional}, the normalization constant
in~\eqref{eq:loi_cond} is indeed computable as it only involves summation over
the $\nt$ themes. As another practical implementation detail, note that the
calculation of~\eqref{eq:loi_cond} can be performed efficiently as the special
function $\Gamma$ (or rather its logarithm) is only ever evaluated at points of
the form $n+\smbeta$ or $n+\nw\smbeta$, where $n$ is an integer, and can thus
be tabulated beforehand.

This formula can readily be used as an alternative decision rule in a
supervised classification setting. We compare this approach with the use of the
naive Bayes classifier in Section~\ref{ssec:bayesiansupervised} below.
Equation~\eqref{eq:loi_cond} is also useful in the context of unsupervised
clustering where it provides the basis for simulation-based inference
procedures to be examined in Section~\ref{ssec:gibbs}.

\section{Supervised Inference}
\label{ssec:bayesiansupervised}

From a Bayesian perspective, the discriminative rule \eqref{eq:loi_cond}
is more principled than the ``naive Bayes'' strategy
\eqref{eq:naivebayes} usually adopted in supervised text clustering.
In this section, we experimentally compare these two approaches.

In \eqref{eq:loi_cond}, $C_{-d}$ is the set of documents whose label is known
and $C_d$ is a particular unlabeled document. To allow for easier comparison
with the naive Bayes classification rule in \eqref{eq:naivebayes}, we rather
denote by $C$ the training corpus, $T$ the associated labels and $C^\star_d$
the test (unlabeled) document. With these notations, \eqref{eq:loi_cond}
becomes
\begin{equation}
\PP(T_d^\star = t |C_d^\star, C, T)
 \propto \left(S_{t} + \smalpha \right)
\frac{\prod_{w=1}^\nw \Gamma(K_{wt}+ C^\star_{wd} +\smbeta)}{\prod_{w=1}^\nw
  \Gamma(K_{wt}+\smbeta)} \frac{\Gamma\left[\sum_{w=1}^{\nw} (K_{wt} +
    \smbeta)\right]}{\Gamma\left[\sum_{w=1}^\nw (K_{wt} + C^\star_{wd}+ \smbeta)\right]}
   \label{eq:loi_predict}
\end{equation}
Comparing with \eqref{eq:naivebayes} we get, after simplification of the Gamma functions,
\[ \left\{ \begin{array}{ll}
           (S_t+\smalpha-1) \frac{\prod_{w=1}^\nw
    (K_{wt}+\smbeta-1)^{C_{wd}^\star}}{(K_t+\nw(\smbeta-1))^{l_{d}^\star}} & \mbox{naive Bayes};\\
        (S_t+\smalpha)\frac
  {\prod_{w=1}^{\nw}\prod_{i=0}^{C_{wd}^\star-1}(K_{wt}+\smbeta+i)}{\prod_{i=0}^{l_d^\star-1}(K_{t}+\nw\smbeta+i)} & \mbox{fully Bayesian approach}.\end{array} \right. \]

If we ignore the offset difference on the hyperparameters (due to the
non coincidence of the mode and the expectation of the multinomial
distribution), note that the two formulas are approximately equivalent
if:
\begin{itemize}
\item[(i)] All counts are $0$ or $1$, hence, $\prod_{i=0}^{C_{wd}^\star-1}(K_{wt}+\smbeta+i)$ simplifies to $(K_{wt}+\smbeta)^{C_{wd}^\star}$.
\item[(ii)] The length $l^\star_d$ of the document is negligible with respect to
  $K_{t}+\nw\smbeta$, therefore, $\prod_{i=0}^{l_d-1}(K_t+\nw\smbeta+i)\approx(K_t+\nw\smbeta)^{l_d^\star}$.
\end{itemize}

\begin{figure}[hbt]
 \centerline{\includegraphics[width=0.8\textwidth]{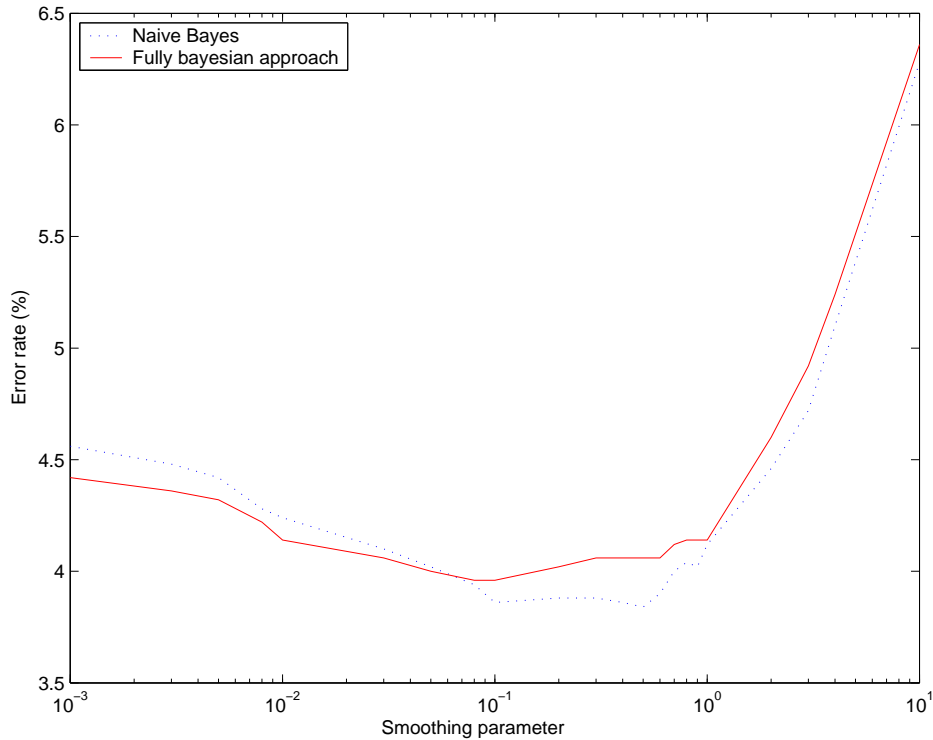}}
\caption{Error rate as a function of $\lambda$.}
\label{fig:supervised}
\end{figure}

To assess the actual difference in performance, we selected 5,000 texts from
the 2000 Reuters Corpus \citep{reuters00corpus}, from five well-defined
categories (arts, sports, health, disasters, employment). In a pre-processing
step, we discard non alphabetic characters such as punctuation, figures, dates
and symbols. For the time being, all words found in the training data are taken
into account.
Words that only occur in the test
corpus are simply ignored. All experiments are performed using ten-fold
cross-validation (with 10 random splits of the corpus). To obtain comparable results, we set:
\[\left\{ \begin{array}{ll}
    \smalpha-1 = 1 \mbox{ and } \smbeta-1=\lambda& \mbox{in the
      Naive Bayes case}\\
    \smalpha = 1 \mbox{ and } \smbeta=\lambda & \mbox{in the
      other case}\end{array} \right., \] 
and change the value of $\lambda$. Figure \ref{fig:supervised} reports the
evolution of the error rate for both classification rules as a function of $\lambda$.

Both approaches yield very comparable results. Naive Bayes
seems to outperform the fully Bayesian approach for larger values of
the smoothing parameter while the converse is true for smaller values.
However, the performance is very similar, since the largest
difference between the scores of both approaches is around $0.2\%$,
corresponding to one text only in our test corpus composed by 500
documents for each fold.

We also tested on other common text classification benchmarks such as
20-Newsgroups \citep{20ngcorpus} and Spam Assassin \citep{spamassassincorpus},
and tried to change the number of documents or size of vocabulary and the
difference never gets statistically significant. Given that the naive Bayes
classifier is known to perform worse than state-of-the-art classification
methods \citep{yang99reexamination,sebastiani02tc}, the fully Bayesian
classifier does not seem to be promising for supervised text classification
tasks.  This is not surprising as the conditions (i) and (ii) discussed above
are nearly satisfied in this context.

We now turn to the unsupervised clustering case, where the fully Bayesian
perspective will prove more useful.

\section{Unsupervised Inference}
\label{sec:unsupervised}

When document labels are unknown, the multinomial mixture model may be used to
create a probabilistic clustering rule. In this context, the performance of the
method is more difficult to assess. We therefore start this section with a
discussion of our evaluation protocol (Section~\ref{ssec:framework}). For
estimating the parameters of the model, we first consider the most widespread
approach, which is based on the use of the \emph{Expectation-Maximization (EM)}
algorithm. It turns out that in the context of large scale text processing
applications, this basic approach is plagued by an acute sensitivity to
initialization conditions. We then consider alternative estimation procedures,
based either on heuristic considerations aimed at reducing the variability of the
EM estimates (Section~\ref{ssec:improving_em}) or on the use of various forms of
Markov chain Monte Carlo simulations (Section~\ref{ssec:gibbs}) and show that
these techniques can yield less variable estimates.

\subsection{Experimental Framework}
\label{ssec:framework}

We will use the same fraction of the Reuters corpus as in Section
\ref{ssec:bayesiansupervised}. As will be discussed below, initialization of the
EM algorithm does play a very important role in obtaining meaningful document
clusters. To evaluate the performance of the model, one option is to look at the
value of the log-likelihood at the end of the learning procedure. However, this
quantity is only available on the training data and does not tell us anything
about the generalization abilities of the model. A more meaningful measure,
commonly used in text applications, is the {\em perplexity}. Its expression on
the test data is:

\[
\widehat{\mathcal{P}}^\star = \exp
[-\frac{1}{l^{\star}}\sum_{d=1}^{n^{\star}_D}\log(\sum_{t=1}^{\nt}\alpha_t\prod_{w=1}^{\nw}\beta_{wt}^{C_{wd}^{\star}})]
\,.\] It quantifies the ability of the model to predict
new documents.
The normalization by the total number of word occurrences $l^{\star}$ in the
test corpus $C^\star$ is conventional and used to allow comparison with simpler
probabilistic models, such as the unigram model, which ignores the document
level. For the sake of coherence, we will also compute perplexity, rather than
log-likelihood, on the training data: their variations are in fact equivalent as
they are identical up to the normalization constant and the exponential
function.

A second indicator, also computable on the training and test data, is obtained
by comparing the cooccurrences of documents between ``equivalent'' clusters in
two clusterings. To do so, we must have a way to establish the best mapping
between clusters in two different clusterings. Provided that the two clusterings
have the same size, this can be done with the so-called Hungarian method
\citep{kuhn55hungarian,frank04hungarian}, an algorithm for computing the best
weighted matching in a bi-partite graph. The complexity of this algorithm is
cubic in the number of clusters involved. Once a one-to-one mapping between
clusters is established, the score we consider is the ratio of documents for
which the two clusterings ``agree'', that
is, which lie into clusters that are mapped by the Hungarian method.
\citep{lange04stability} describes in more detail how this method can be used to
evaluate clustering algorithms.

A limitation of the evaluation with the Hungarian method is that it is not
suited to compare two soft clusterings with different number of classes and
especially the cases where one class in clustering $A$ is split into two classes
in clustering $B$. There exist other information-based measures built, such as
the \emph{Relative Information Gain}, that do not suffer from this limitation
but present other drawbacks, such as undesirable behaviors for distributions
close to equiprobability. We do not consider those here, as cooccurrence scores
obtained with the Hungarian method are easier to interpret (see 
\citealp{rigouste05evaluation}, for results on the same database quantified in
terms of mutual information).

\subsection{Expectation-Maximization algorithm}
\label{ssec:em}

In an unsupervised setting, the \emph{maximum a posteriori} estimates are
obtained by maximizing the posterior distribution given in \eqref{eq:paramdist}.
The resulting maximization program is unfortunately not tractable. It is
however possible to devise an iterative estimation procedure, based on the
Expecta\-tion-Maximization (EM) algorithm. Denoting respectively by $\alpha'$
and $\beta'$ the current estimates of the para\-meters and by $T_d$ the latent
(unobservable) theme of document $d$, it is straightforward to check that each
iteration of the EM algorithm updates the parameters according to:
\begin{align}
&\PP(T_d=t|C;\alpha',\beta')  = \frac{\alpha_t'\prod_{w=1}^{\nw}\beta_{wt}'^{C_{wd}}}{\sum_{t'=1}^{\nt}\alpha_{t'}'\prod_{w=1}^{\nw}\beta_{wt'}'^{C_{wd}}} \label{eq:EM:postprob}\\
&\alpha_t  \propto \smalpha - 1 + \sum_{d=1}^{\nd}\PP(T_d=t|C;\alpha',\beta') \label{eq:EM:alpha} \\
&\beta_{wt}  \propto \smbeta -1 + \sum_{d=1}^{\nd}C_{wd}\PP(T_d=t|C;\alpha',\beta') \label{eq:EM:beta}
\end{align}
where the normalization factors are determined by the constraints:
\[
\left\{ \begin{array}{ll}
  \sum_{t=1}^{\nt} \alpha_t = 1&\\
  \sum_{w=1}^{\nw} \beta_{wt} =1& \mbox{for $t$ in
    $\{1,\dots,\nt\}$}.
\end{array} \right. 
\]

In the remainder of this section, we present the results of a series of
experiments based on the use of the EM algorithm. We first discuss issues
related to the initialization strategy, before empirically studying the
influence of the smoothing parameters. The main findings of these experiments is
that the EM estimates are very unstable and vary greatly depending on the
initial conditions: this outlines a limitation of the EM algorithm, i.e. its
difficulty to cope with the very high number of local maxima in high dimensional
spaces. In comparison, the influence of the smoothing parameter is moderate, and
its tuning should not be considered a major issue.

\subsubsection{Initialization}

It is important to realize that the EM algorithm allows to go back and forth
between the values of the parameters $\alpha$ and $\beta$ and the values of the
posterior probabilities $\PP(T_d=t|C; \alpha,\beta)$ using formulas
\eqref{eq:EM:postprob}, \eqref{eq:EM:alpha} and \eqref{eq:EM:beta}. Therefore,
the EM algorithm can be initialized either from the E-step, providing initial
values for the parameters, or from the M-step, providing initial values for
the posterior probabilities (that is, roughly speaking, an initial soft
clustering). There are various reasons to prefer the second solution:
\begin{itemize}
\item Initializing $\beta$ requires to come up with a reasonable value for a
  very large number of parameters. A random initialization scheme is out of the
  question, as it almost always yields very unrealistic values in the parameter
  space; an alternative would be to consider small deviations from the unigram
  word frequencies: it is however unclear how large these deviations should be.
\item Initializing on posterior probabilities can be done without any knowledge
  of the model: for instance, it can be performed without knowing the vocabulary
  size. Section \ref{ssec:improving_em} will show why this is a desirable
  property.
\end{itemize}
Consequently, in the rest of this article, we will only consider initialization
schemes that are based on the posterior theme probabilities associated with each
document. A good option is to make sure that, initially, all clusters
significantly overlap. Our ``Dirichlet'' initialization consists in sampling,
independently for each document, an initial (fictitious) configuration of
posterior probabilities from an exchangeable Dirichlet distribution. In
practice, we used the uniform distribution over the $\nt$-dimensional
probability simplex (Dirichlet with parameter 1). As the EM iterations tend to
amplify even the smaller discrepancies between the components, the variability
of the final estimates was not significantly reduced when initializing from
exchangeable Dirichlet distributions with lower variance (ie., higher parameter
value).

\begin{figure}[hbt]
 \centerline{\includegraphics[width=0.8\textwidth]{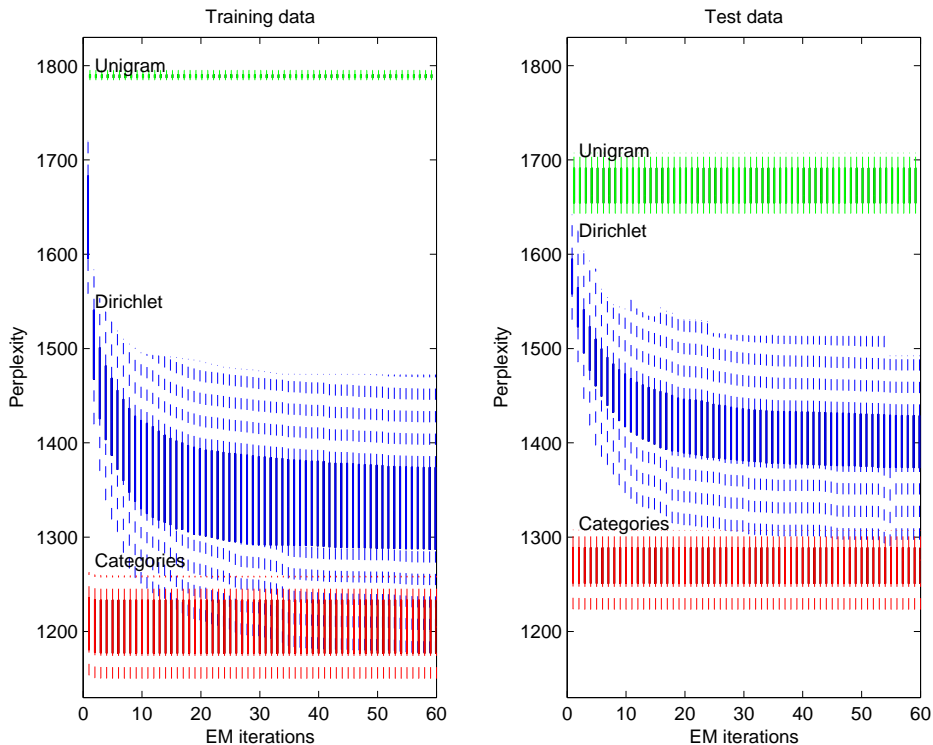}}
\caption{Evolution of Perplexity on training and test data over the EM iterations.}
\label{fig:boxplotperp}
\end{figure}

To get an idea about the best achievable performance, we also used the Reuters
categories as initialization. We establish a one-to-one mapping between the
mixture components and the Reuters categories, setting for each document the
initial posterior probability in~\eqref{eq:EM:postprob} to 1 for the corresponding theme.
Figure \ref{fig:boxplotperp} displays the corresponding perplexity on the
training and test sets as a function of the number of iterations. Results
are averaged over 10 folds and 30 initializations per fold and are represented
with box-and-whisker curves: the boxes being drawn between the lower and upper
quartiles, the whiskers extending down and up to $\pm 1.5$ times the
interquartile range (the outliers, a couple of runs out of the 300, have been
removed).

The variations are quite similar on both (training and test) datasets. The main
difference is that test perplexity scores are worse than training perplexity
scores. This classical phenomenon is an instance of overfitting. Due to the way
the indexing vocabulary is selected (discarding words that do not occur in the
training data), this effect is not observed for the unigram model\footnote{For both models, the fit is better on the training set than on the test set,
  which should be reflected by an increase in perplexity from one dataset to the
  other. However, as we ignore those (rare) words which only appear in the test
  data, the average probability of the remaining words in this corpus is
  somewhat artificially increased. For the unigram model, which is less prone to
  overfitting, this effect is the strongest, yielding a quite unexpected overall
  improvement of perplexity from the training to the test set.}.

The most striking observation is that the gap between both initialization
strategies is huge. With the Dirichlet initialization, we are able to predict
the word distribution more accurately than with the unigram model but much worse
than with the somewhat ideal initialization. This gap is also patent for the
cooccurrence scores with a final ratio of 0.95 for the ``Reuters categories'' initialization
and an average around 0.6 for the Dirichlet initialization on test data.

\begin{figure}[hbt]
 \centerline{\includegraphics[width=0.8\textwidth]{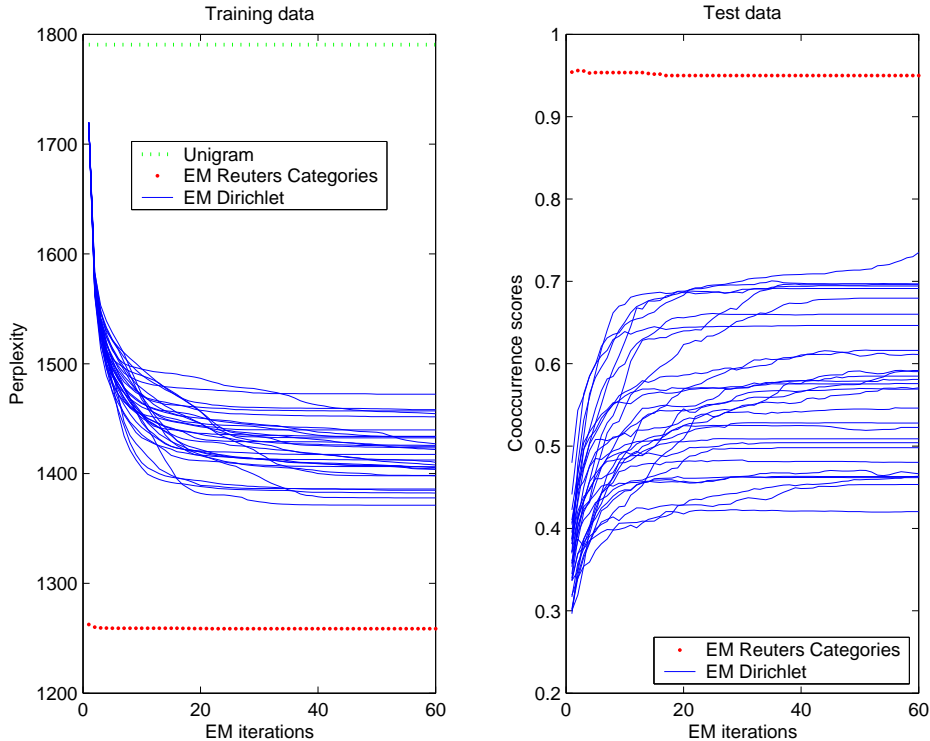}}
 \caption{Evolution of Perplexity and Cooccurrence scores over the EM
   iterations for different Dirichlet initializations for the first
   fold.}
\label{fig:ex_em}
\end{figure}

Given that the Dirichlet initialization involves random sampling, it is worth
checking how the performance change from one run to another. We report in Figure
\ref{fig:ex_em} the values of training perplexity and test cooccurrence scores
for various runs on the first fold\footnote{In the rest of this article, perplexity measurements are only performed on the
  training data, for test data, we use the cooccurrence score as our main
  evaluation measure. Depending on the readability of the results, we either
  plot all runs, as in Figure \ref{fig:ex_em}, or a box-and-whisker curve, as in
  Figure \ref{fig:boxplotperp}.}.
As can be seen more clearly on this figure, the variability from one
initialization to another is very high for both measures: for instance, the
cooccurrence score varies from about $0.4$ to more than $0.7$. This variability
is a symptom of the inability of the EM algorithm to avoid being
trapped in one of the abundant local maxima which exist in the high-dimensional
parameter space.

\subsubsection{Influence of the smoothing parameter}

\begin{figure}[hbt]
  \centerline{ \includegraphics[width=0.8\textwidth]{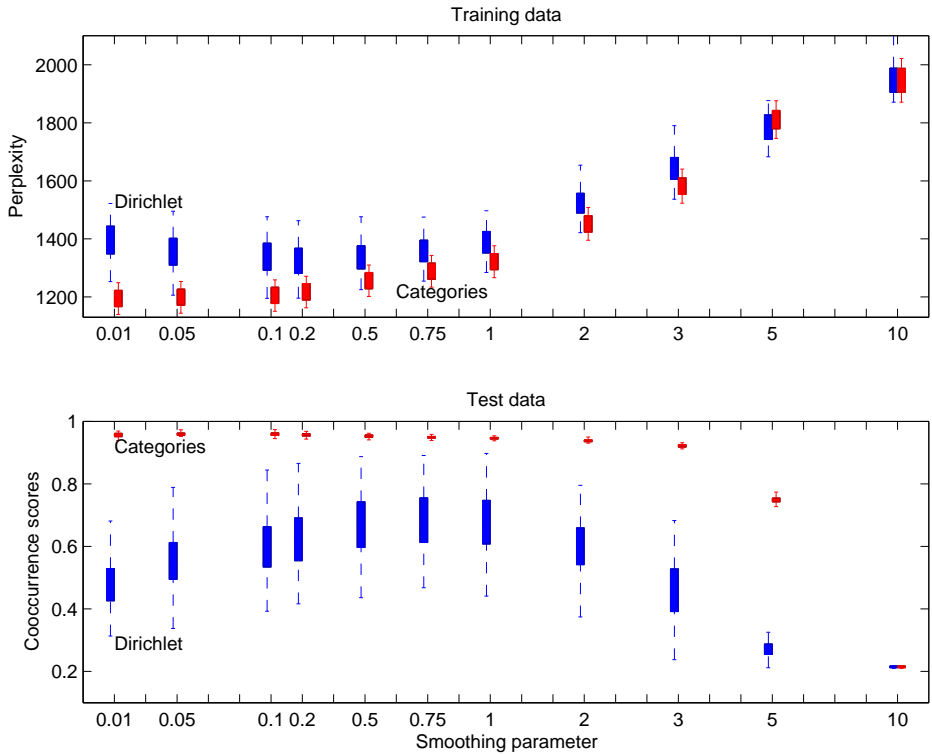}}
  \caption{Evolution of training perplexity and test cooccurrence
    scores over the smoothing parameter $\smbeta-1$.}
  \label{fig:perp_cooc_smoothing}
\end{figure}

Figure~\ref{fig:perp_cooc_smoothing} depicts the influence of the smoothing
parameter $\smbeta-1$ in terms of perplexity and cooccurrence scores. We do not consider here the influence of $\smalpha-1$, which is, in our
  context, always negligible with respect to the sum over documents of the
  themes posterior probabilities.
For the Reuters categories initialization, there is almost no difference in perplexity
scores for small values of $\smbeta - 1$ (i.e. when $\smbeta-1\leq0.2$). The
performance degrades steadily for larger values, showing that some information is
lost, probably in the set of rare words (since smoothing primarily concerns
parameters corresponding to very few occurrences). Similarly, for the
Dirichlet initialization, the variations in perplexity are moderate for
smoothing values in the range $0.01$ to $1$, yet there is a more distinguishable
optimum, around $0.2$. Using some prior information about the fact that word
probabilities should not get too small helps to fit the distribution of new
data, even for words that are rarely (or even never) seen in association with a
given theme.

These observations are confirmed by the observation of the test cooccurrence
scores. First, except when using very large ($5$ or more) values of the
smoothing parameters, which yields a serious drop in performance, the
categorization accuracy is rather insensitive to the smoothing parameter for the
Reuters categories initialization. Of more practical interest however is the
behavior for the Dirichlet initialization: the variations in performance are
again moderate, with however a higher optimum value around $0.75$. A possible
explanation of this observation that more smoothing improves categorization
capabilities (even if it slightly degrades distribution fit) is that the model
is so coarse and the data so sparse that only quite frequent words are helpful
in categorizing; the other words are essentially misleading, unless properly
initialized. This suggests that removing rare words from the vocabulary could
improve the classification accuracy.

All in all, changing the values of $\smalpha-1$ and $\smbeta-1$ does
not make the most important differences in the results, as long as
they remain within reasonable bounds. Thus, in the rest of this article,
we set them respectively to $0$ and $0.1$.

\subsection{EM and deterministic clustering}
\label{ssec:equivalence_with_kmeans}

A somewhat unexpected property of the multinomial mixture model is that a huge
fraction of posterior probabilities (that a document belongs to a given theme)
is in fact very close to $0$ or $1$. Indeed, when starting from the Reuters
categories, the proportion of texts classified in only one given theme (that is,
with probability one, up to machine precision) is almost 100\%. As we start from
the opposite point of ``extreme fuzziness'', this effect is not as strong with
the Dirichlet initialization. Nevertheless, after the fifth iteration, more
than 90\% of the documents are categorized with almost absolute certainty. This
suggests that in the context of large-dimensional textual databases, the
multinomial mixture model in fact behaves like a deterministic clustering
algorithm.

This intuition has been experimentally confirmed as follows, implementing a ``hard''
(deterministic) clustering version of the EM algorithm, in which the E-step
uses deterministic rather than probabilistic theme assignments. This
algorithm can be seen as an instance of a $K$-means algorithm, where the
similarity between a text $d\in\{1,\ldots,n_D\}$ and theme (or cluster)
$t\in\{1,\ldots,\nt\}$ is computed as:
\[
\operatorname{dist}(d,t) = - \sum_{w=1}^{\nw} C_{wd} \log( \beta_{wt}) + \log \alpha_t
\]
Up to a constant term, which only depends on the document, the first term is the
Bregman divergence \citep{banerjee05bregman} between a theme specific
distribution and the document, viewed as an empirical probability distribution
over words. This measure is computed for every document and every theme, and
each document is assigned to the closest theme. The reestimation of the
parameters $\beta_{wt}$ is still performed according to~\eqref{eq:EM:beta},
where the posterior ``probabilities'' are either $0$ or $1$. The weight $\alpha_t$ simply
becomes the proportion of documents in theme $t$ and $\beta_{wt}$ the ratio of
the number of occurrences of $w$ in theme $t$ over the total number of
occurrences in documents in theme $t$.

\begin{eqnarray*}
  \alpha_t&=&\frac{\#\{d: T_d = t\}}{n_D}\\
  \beta_{wt}&=&\frac{\sum_{\{d: T_d = t\}}C_{wd}}{\sum_{w=1}^{n_W}\sum_{\{d: T_d = t\}}C_{wd}}
\end{eqnarray*}

This algorithm was applied to the same dataset, with the same initialization
procedures as above. At the end of each iteration, we compute the cooccurrence
score between the probabilistic clustering produced by EM and the hard clustering
produced by this version of K-means. 
\begin{itemize}
\item With the Reuters Categories initialization, the cooccurrence
  score between both clusterings is $1$ after one iteration.
\item With the Dirichlet initialization, the score between the soft and hard
  clustering quickly converges to $1$ and is greater than $0.99$ after five
  iterations.
\end{itemize}
In both cases, the outputs of the probabilistic and hard methods become indiscernible
after very few iterations. We believe that this behavior of EM can be partly
explained by the large dimensionality of the space of documents\footnote{The
  vocabulary contains more than 40,000 words.}. This assumption has been
verified with experiments on artificially simulated datasets, which are not reported here for reason of space.

\subsection{Improving EM via dimensionality reduction}
\label{ssec:improving_em}

In this section, we push further our intuition that removing rare words should
improve the performance of the EM algorithm and should alleviate the
variability phenomenons observed in the previous section. After studying the
effect of dimensionality reduction, we propose a novel strategy based on
iterative inference.

\subsubsection{Adjusting the Vocabulary Size}
\label{sssec:vocsize}

\begin{figure}[hbt]
 \centerline{\includegraphics[width=0.8\textwidth]{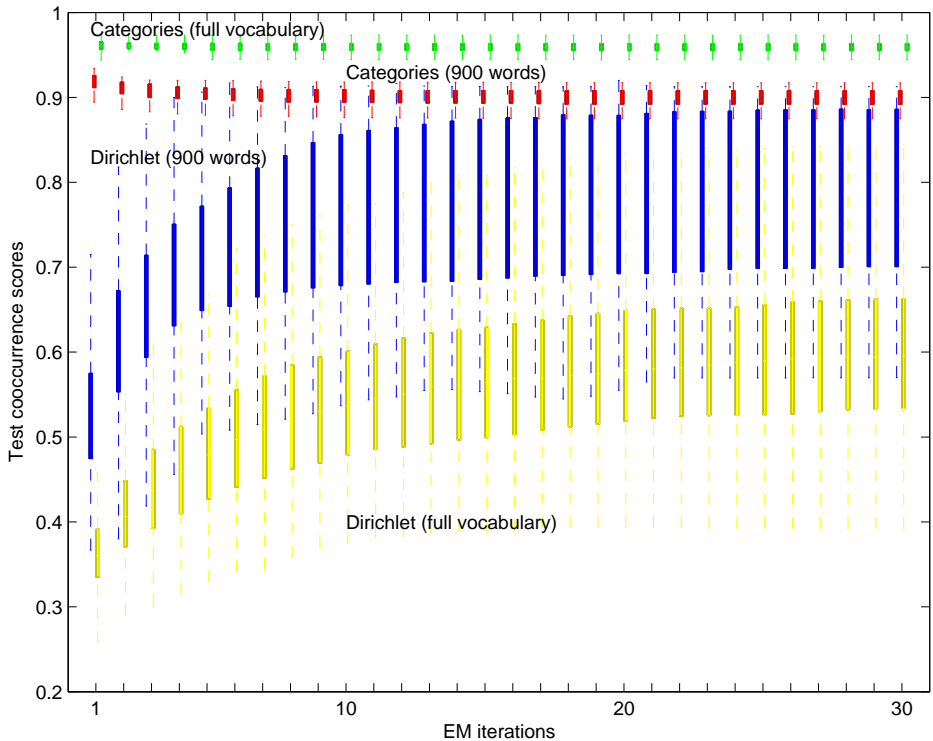}}
\caption{Evolution of test cooccurrence scores over the EM
  iterations with a vocabulary of size 900.}
\label{fig:em_900words}
\end{figure}

Having decided to ignore part of the vocabulary, the next question is whether we
should rather discard the rare words or the frequent words. In this section, we
experimentally assess these strategies, by removing consecutively tens, hundreds
and thousands of terms from the indexing vocabulary. The words that are
discarded are simply removed from the count matrix\footnote{%
  An alternative option, that we do not consider here, would be to replace all
  the words that do not appear in the vocabulary by a generic
  ``out-of-vocabulary'' token. The main reason for not using this trick is that
  this generic token tends to receive a non-negligible probability mass; as a
  consequence, documents containing several unknown words tend to look more
  similar than they really are.}%
. Results presented in Figure~\ref{fig:em_900words} suggest that the performance
of the model with the Dirichlet initialization can be substantially improved
by keeping a limited number of frequent words (900 out of 40,000).

\begin{figure}[hbt]
 \centerline{\includegraphics[width=0.8\textwidth]{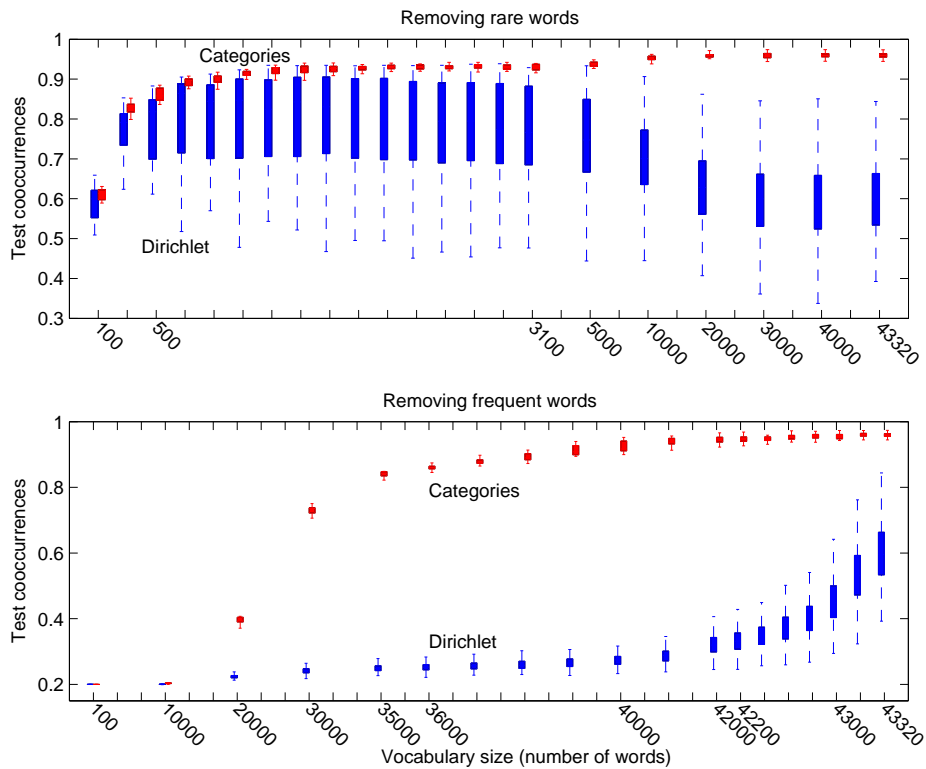}}
\caption{Evolution of test cooccurrence scores over the size of
  vocabulary with two different strategies: discarding most rare words
or discarding most frequent words.}
\label{fig:testcooc_change_voc_size}
\end{figure}

When varying the size of the vocabulary, perplexity measurements are
meaningless, as the reduction of dimensionality has an impact on perplexity
which is hard to distinguish from the variations due to a possible better fit of
the model. The test cooccurrence score, on the other hand, is meaningful even
when with varying vocabulary sizes. Figure~\ref{fig:testcooc_change_voc_size}
plots the test cooccurrence scores at the end of the 30th EM iteration as a
function of the vocabulary size. For the sake of readability, the scale of the
x-axis is not regular but rather focuses on the interesting parts: the interval
between $100$ and $3,000$ words, which corresponds to keeping only the frequent
words, and the region above $40,000$ (out of a complete vocabulary of $43,320$
forms), which corresponds to keeping only the rare words. This choice is
motivated by the well-known fact that most of the occurrences (and therefore
most of the information) are due to the most frequent words: for instance, the
$3,320$ most frequent words account for about $75\%$ of the total number of
occurrences.

The upper graph in Figure~\ref{fig:testcooc_change_voc_size} shows that removing
rare words always hurts when using the Reuters categories initialization. In
contrast, with the Dirichlet initialization, considering a reduced vocabulary
(between $300$ and $3,000$ words) clearly improves the performance. The somewhat
optimal size of the vocabulary, as far as this specific measure is concerned,
seems to be around $1,000$. Also importantly, the performance seems much more
stable when using reduced versions of the vocabulary, an effect we did not
manage to achieve by adjusting the smoothing parameter. We will come back to
this in the next section. It suffices to say here that the best score obtained
with the Dirichlet initialization is still far behind the performance attained
with the Reuters categories initialization. This agrees with our previous observation that
even the rarest word are informative, when properly initialized.

Less surprisingly, on the lower portion of the graph, one can see that removing
the frequent words almost always hurts the performance. It is only in the case
of the Reuters categories initialization that the removal of the $100$ most frequent
words actually yields a slight improvement of performance. Then the score
steadily decreases with the removal of frequent words. The score is almost
$0.2$ (random agreement) with $20,000$ rare words, which is not surprising, as,
in this case, the vocabulary mainly contains words occurring only once (so-called
\textit{hapax legomena}) in the corpus, reducing texts to at most a dozen of
terms.

To sum-up, there are two important lessons to draw from these experiments:
\begin{itemize}
\item Reducing the dimensionality (vocabulary size) while the number of examples
  (size of the corpus) remains the same helps the inference procedure;
\item Using a reduced vocabulary allows to significantly reduce the variability of
  the parameter estimates.
\end{itemize}
We now consider ways to use these remarks to improve our estimation procedure.

\subsubsection {Iterative estimation procedures}
\label{sssec:iterative}

In this section, we look for ways to reduce the variability of the
clustering: our main goal here being that an end-user should get
sufficiently reliable results without having to run the program
several times and/or to worry about evaluation measures.

\paragraph{Incremental vocabulary}

The first idea is to take advantage of our previous observations that reducing
the dimension of the problem seems to make the EM algorithm less dependent on
initial conditions. This suggests to obtain robust posterior probabilities using
a reduced vocabulary, and to use them for initializing new rounds of EM
iterations, with a larger vocabulary. Proceeding this way allows us to
circumvent the problem of initializing the $\beta$ parameters corresponding to
rare words, as we start from the other step of the algorithm (the M-step).
When the vocabulary size is increased, the probabilities associated with new
words are implicitly initialized on their average count in the corpus, weighted
by the current posterior probabilities.This iterative procedure has the net
effect of decomposing the inference process into several steps, each being
sufficiently stable to yield estimates having both a small variance and good
generalization performance.

\begin{figure}[hbt]
\centerline{\includegraphics[width=0.8\textwidth]{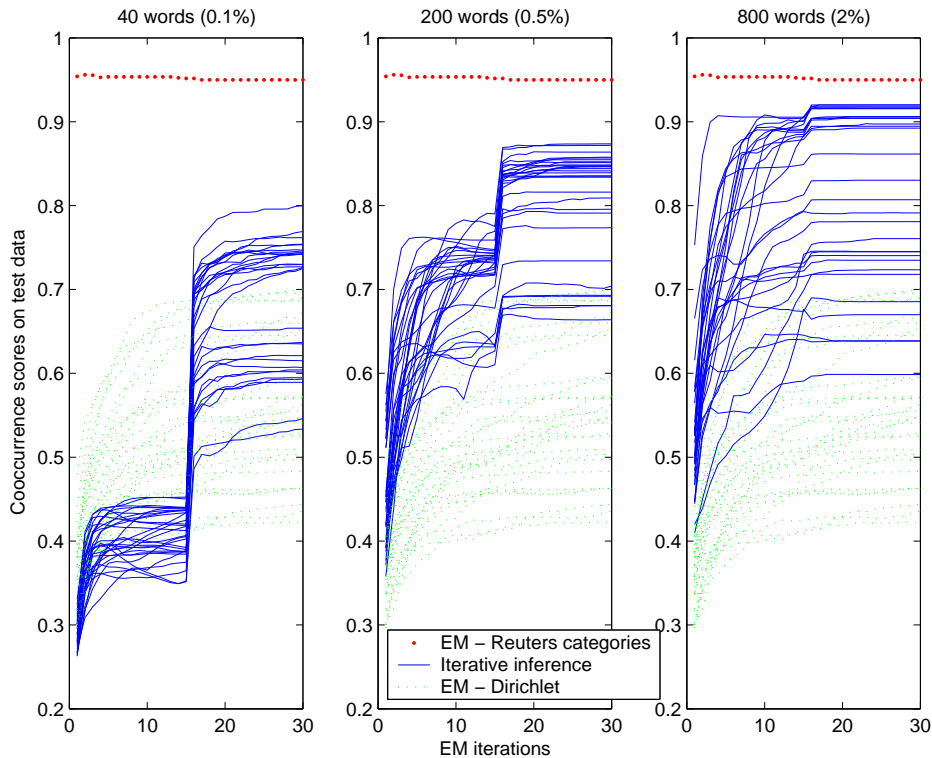}}
\caption{Evolution of cooccurrence scores on test data with different
  vocabulary sizes. The first 15 iterations are conducted on reduced
  vocabularies (the size of which is reported on top the
  figures), while the last 15 are use the complete vocabulary.}
\label{fig:iterative}
\end{figure}

Figure~\ref{fig:iterative} displays the results of the following set of
experiments: we perform 15 EM iterations with a reduced vocabulary, save the
values of posterior probabilities at the end of the 15th iteration, and use
these values to initialize another round of 15 EM iterations, using the full
vocabulary. Our earlier results obtained using the full vocabulary are also
reported for comparison. The influence of the initial vocabulary size is
important: as it is increased, the maximal score gets somewhat better but the
results are more variable.

\begin{figure}[hbt]
\centerline{\includegraphics[width=0.8\textwidth]{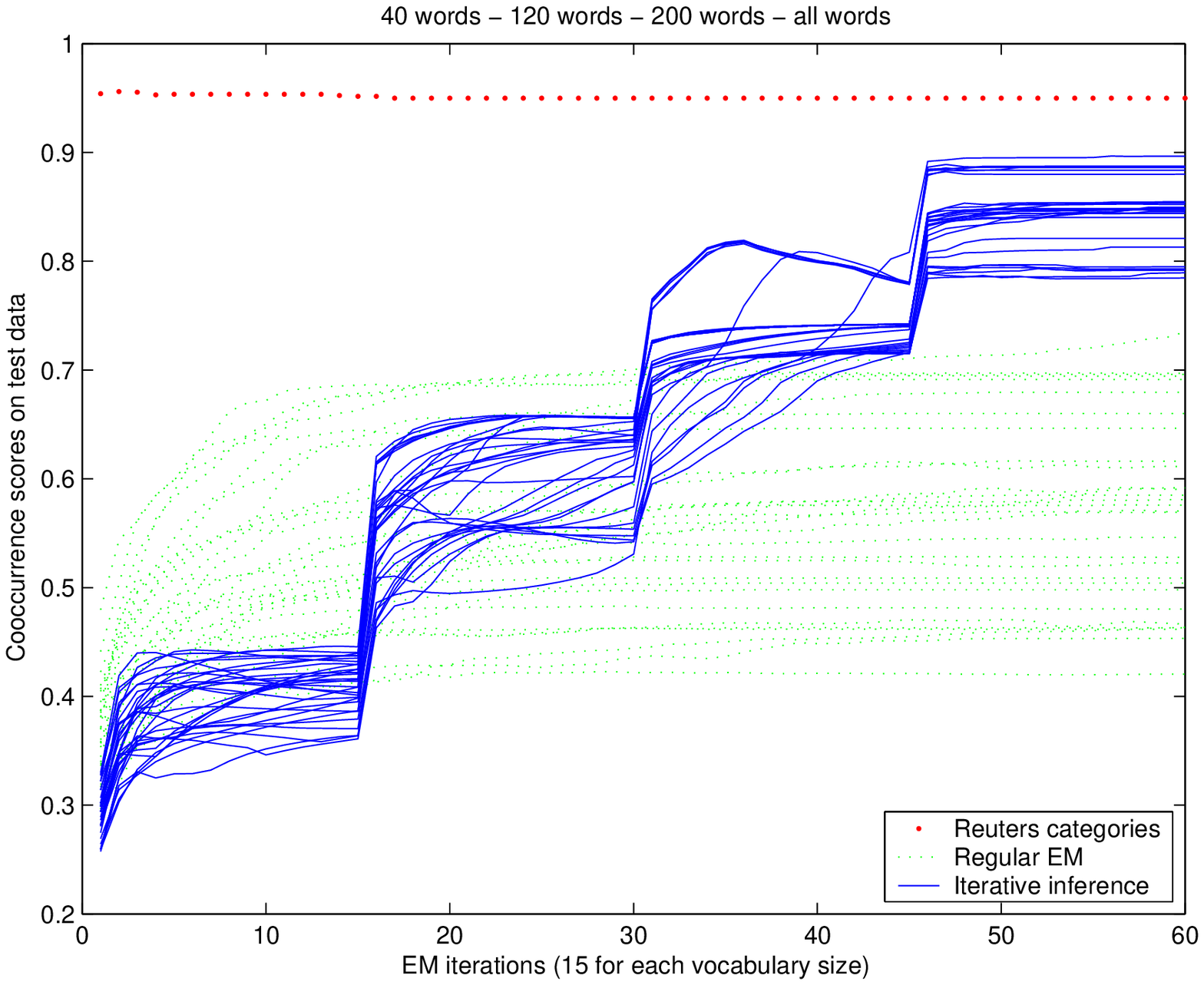}}
\caption{Evolution of cooccurrence scores over the different steps of
  an iterative algorithm (30 runs on the same fold).}
\label{fig:iterative_4steps}
\end{figure}

These results can be improved by making the estimation process more gradual,
thus reducing the variability of our estimates. Such experiments are reported in
Figure~\ref{fig:iterative_4steps} where we use four different steps. Proceeding
this way, both the maximal and minimal scores lie within an acceptable range of
performance. It is clear from these experiments that the choice of the
successive sizes of vocabulary is particularly difficult, being a tight
compromise between quality and stability. It remains to be seen how to devise a
principled approach for finding such appropriate vocabulary increments.

\paragraph{Multiple restarts}

\begin{figure}[hbt]
\centerline{\includegraphics[width=0.8\textwidth]{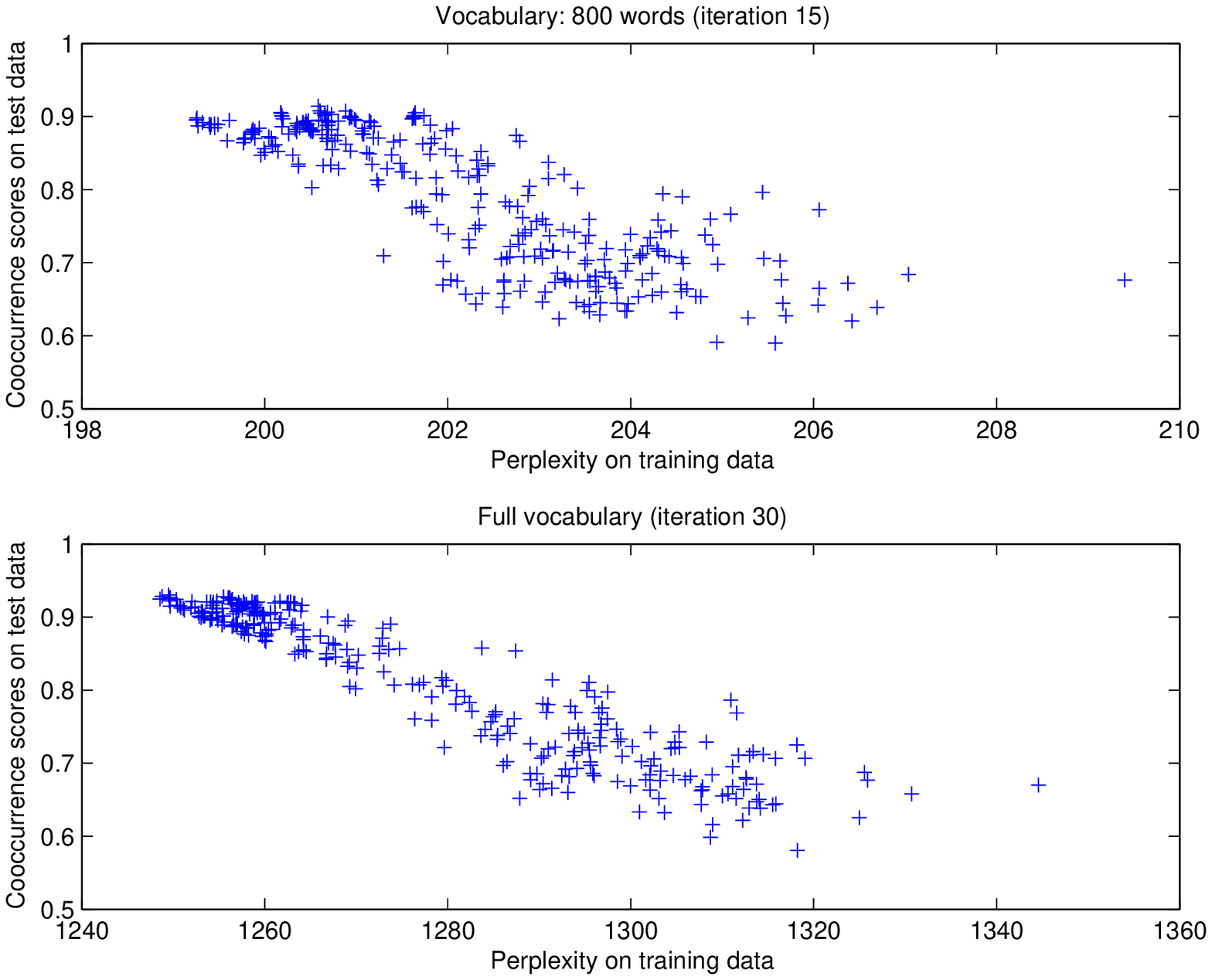}}
\caption{Correlation between training perplexity and test cooccurrence
  scores.}
\label{fig:correlation_perp_cooc}
\end{figure}

Another usual approach in optimization problems where the large number of local
optima yields unstable results is to perform \emph{multiple restarts}
and pick up the best run according to some criterion. From this point of view, a
sensible strategy is to choose the vocabulary size yielding the best maximum
performance (for instance, Figure~\ref{fig:iterative} suggests that starting
with $800$ words is a reasonable choice), run several trials and select the
parameter set yielding the best cooccurrence score on the test data. For lack of
this information (as would be the case in a real-life study, where no reference
clustering is available), a legitimate question to ask is whether the training
perplexity could be used instead as a reliable indicator of the quality of
parameter settings. The answer is positive, as is shown in
Figure~\ref{fig:correlation_perp_cooc}.

This figure reports results of the following experiments: after 15 EM iterations
using a reduced vocabulary of $800$ words, we consider the complete vocabulary for
another 15 additional EM iterations. Training set perplexity is computed at the
end of iteration 15 and at the end of iteration 30. These measurements are
repeated 30 times for each of the 10 folds. The test cooccurrence scores
(somewhat representing the quality of clustering) are plotted as a function
of this training perplexity in Figure \ref{fig:correlation_perp_cooc}. There is
a clear inverse correlation, especially in the area of the best runs (low
perplexity values--large cooccurrence scores) we are interested in. Selecting
the run with lowest perplexity yields acceptable performance in both cases and
the correlation is even stronger on the full vocabulary (it is therefore worth
performing 15 more iterations with all the words).

In summary, we have presented in this section two inference strategies which
significantly improve over a basic implementation of the EM algorithm:
\begin{itemize}
\item split the vocabulary in several bins (at least 4) based on frequency; run
  EM on the smallest set and iteratively add words and rerun EM.
\item discard rare words, run several rounds of EM iterations, keep the run
  yielding the best training perplexity.
\end{itemize}

\citep{rigouste05ssp} reports experiments which show that these strategies can
be combined, yielding improved estimates of the parameters.

\subsection{Gibbs sampling algorithm}
\label{ssec:gibbs}

In this section, we experiment with an alternative inference method, Gibbs
sampling. The first subsection presents the results obtained with the most
``naive'' Gibbs sampling algorithm, which is then compared with a
\emph{Rao-Blackwellized} version relying on the integrated formula introduced
in~\eqref{eq:loi_cond}.

\subsubsection {Sampling from the EM formulas}
\label{sssec:gibbs1results}

To apply Gibbs sampling, we first need to identify sets of variables whose
values may be sampled from their joint conditional distribution given the
other variables. In our case, the most
straightforward way to achieve this is to use the EM update equations
\eqref{eq:EM:postprob}, \eqref{eq:EM:alpha}, \eqref{eq:EM:beta}.  Hence, we may
repeatedly:
\begin{itemize}
\item sample a theme indicator in $\{1,\dots,\nt\}$ for each document from a
  multinomial distribution whose parameter is given by the posterior
  probability that the document belongs to each of the themes;
\item sample values for $\alpha,\beta$ which, conditionally upon the theme
  indicators, follow Dirichlet distributions;
\item compute new posterior probabilities according to
  \eqref{eq:EM:postprob}.
\end{itemize}

\begin{figure}[hbt]
 \centerline{\includegraphics[width=0.8\textwidth]{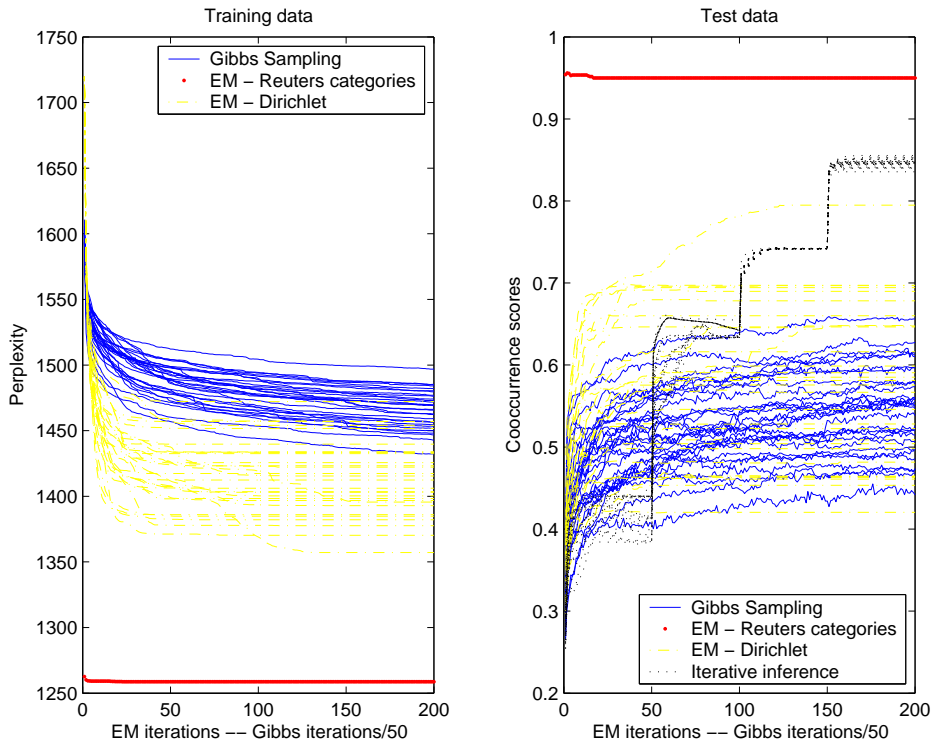}}
\caption{Evolution of perplexity and cooccurrence scores over
  the EM--Gibbs Sampling iterations.}
\label{fig:gibbswithemformulas}
\end{figure}

Figure~\ref{fig:gibbswithemformulas} displays the evolution of the training
perplexity and the test cooccurrence score for 200 runs of the Gibbs sampler
(ran for 10,000 iterations on one fold), compared to the regular EM algorithm
and the iterative inference method described in Section \ref{ssec:improving_em}.
The performance varies greatly from one run to another and, occasionally, large
changes occur during a particular run. This behavior suggests that, in this
context, the Gibbs sampler does not really attain its objective and gets
trapped, like the EM algorithm, in local modes. Hence, one does not really
sample from the actual posterior distribution but rather from the posterior
restricted to a ``small'' subset of the space of latent variables and
parameters. Results in terms of perplexity and cooccurrence scores are in the
same ballpark as those obtained with the EM algorithm, several levels below the
ones obtained with the ad-hoc inference method of Section \ref{sssec:iterative}.

\subsubsection{Rao-Blackwellized Gibbs Sampling}
\label{sec:gibbs2results}

There is actually no need to simulate the parameters $\alpha$ and
$\beta$, as they can be integrated out when considering the
conditional distribution of a single theme given in \eqref{eq:loi_cond}.
We then obtain an estimate of the distribution of the themes $T$
of all documents by applying the Gibbs sampling algorithm to simulate, in
turn, every latent theme $T_d$, conditioned on the theme assignment
of all other documents. This strategy, which aims at reducing the
number of dimensions of the sampling space, is known as
\emph{Rao-Blackwellized sampling}, and often produces good results
\citep{robert99montecarlo}. Note that if the document $d$ is one word
long, this approach is identical to the Gibbs sampling algorithm described in
\citep{griffiths02probabilistic} for the LDA model (using the identity
$\Gamma(a+1) = a \Gamma(a)$).

\begin{figure}[hbt]
  \centering
  \includegraphics[width=0.8\textwidth]{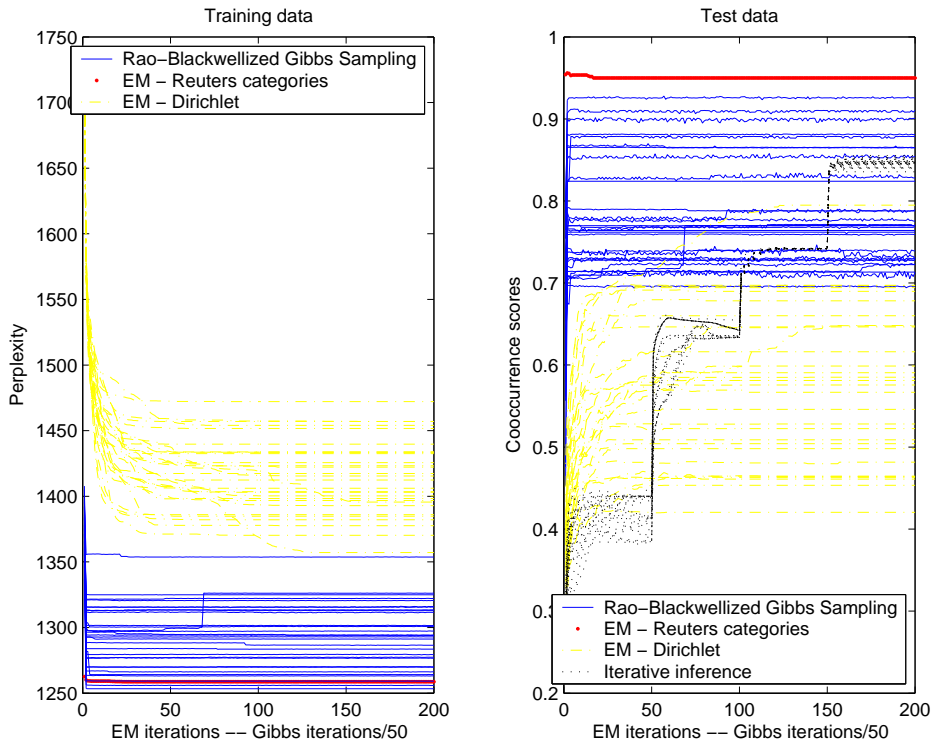}
  \caption{Evolution of the different measures for Rao-Blackwellized Gibbs Sampling}
  \label{fig:smartgibbs}
\end{figure}

Figure \ref{fig:smartgibbs} displays the training perplexity and the
test cooccurrence scores for 30 independent random initializations of
the Gibbs sampler, compared to the same references as in the previous
section. We plot results obtained on 200 samples, each corresponding
to 10,000 complete cycles on one fold. The Gibbs sampler outperforms
the basic EM algorithm for almost all runs. Its performance is in the
same range as the iterative method, albeit much more variable (the
cooccurrence score lies in the range 70\% to 95\%). The sampler
trajectories also suggest that the Gibbs sampler is not really
irreducible in this context and only explores local modes.

This alternative implementation of the Gibbs sampling procedure is
obviously much better than our first, arguably more naive,
attempt: not only does it yield consistently better performance, but
it is also much faster. Thanks to the
tabulation of the Gamma function, the deterministic computations needed
for both versions of the sampler are comparable. But the Gibbs sampler based
on the EM formulas requires generating $\nt+1$ Dirichlet samples (with
respective dimensions $\nw$ and $\nt$) for a rough total of
approximately $\nw\nt$ Gamma distributed variables for the M-step,
and $\nd$ samples from $\nt$-dimensional discrete distributions for
the E-step. In comparison, the Rao-Blackwellized Gibbs sampling
only requires $\nd$ $\nt$-dimensional samples from discrete distributions.
The difference is significant: our C-coded implementation of
the latter algorithm runs 20 times faster than the vanilla Gibbs sampler.

\section{Conclusion}

In this article, we have presented several methods for estimating the
parameters of the multinomial mixture model for text clustering. A
systematic evaluation framework based on various measures allowed us
to understand the discrepancy between the performance typically
obtained with a single run of the EM algorithm and the best scores we
could possibly attain when initializing on a somewhat ideal
clustering.

Based on the intuition that the high dimensionality incurred by a
``bag-of-word'' representation of texts is directly responsible for
this undesirable behavior of the EM algorithm, we have analyzed the
benefits of reducing the size of the vocabulary and suggested a
heuristic inference method which yields a significant improvement in
comparison to the basic application of the EM algorithm.

We have also investigated the use of Gibbs sampling, and proposed two
different approaches. The Rao-Blackwellized version, which takes
advantage of analytic marginalization formulas clearly outperforms the
other, more straightforward, implementation. Performance obtained with
Gibbs sampling are close to the ones obtained with the iterative
inference method, albeit more dependent on initial conditions.

Altogether, these results clearly highlight the too often overlooked
fact that the inference of probabilistic models in high-dimensional
spaces, as is typically required for text mining applications, is
prone to an extreme variability of estimators.

This work is currently extended in several directions. Further
investigations of the multinomial mixture model are certainly
required, notably aiming at (i) analyzing its behavior when used with
very large numbers (several hundreds) number of themes, as in
\citep{blei02latent}; (ii) investigating model selection procedures to
see how they can help discover the proper number of themes; (iii)
reducing the overall complexity of the training:
both the EM-based and the Gibbs
sampling algorithm require to iterate over each document, an
unrealistic requirement for very large databases.

Another promising line of research is to consider alternative models:
the multinomial mixture model can be improved in multiple ways: (i)
its modeling of the count matrix is unsatisfactory, especially as it
does not take in account typical effects of word occurrence
distributions \citep{church95poisson,katz96distribution}: this
suggests to consider alternative, albeit more complex models of the
counts; (ii) the one document-one theme assumption is also
restrictive, pleading for alternative models such as LDA
\citep{blei02latent} or GAP \citep{canny04gap}: preliminary experiments
with the former model however suggest that it might be faced with the
same type of variability issues as the multinomial mixture model \citep{rigouste06lda}.

\bibliographystyle{abbrvnat}
\bibliography{maindocument}

\end{document}